\documentclass[epj]{svjour}
\usepackage{times}
\usepackage{graphicx}
\newcommand{\pd}[2]{\frac{\partial #1}{\partial #2}}
\begin{document} 
\title{On the Microscopic Foundations of Elasticity} 
\author{I. Goldhirsch\inst{1} \and C. Goldenberg\inst{2}} 
\institute{Department of Fluid Mechanics and Heat
  Transfer, Faculty of Engineering, Tel-Aviv University, Ramat-Aviv, Tel-Aviv
  69978, Israel; \email{isaac@eng.tau.ac.il} \and School of Physics and
  Astronomy, Tel-Aviv University, Ramat-Aviv, Tel-Aviv 69978, Israel;
  \email{chayg@post.tau.ac.il}} 
\date{March 18, 2002} 

\abstract{The modeling of the elastic properties of disordered or nanoscale
  solids requires the foundations of the theory of elasticity to be revisited,
  as one explores scales at which this theory may no longer hold. The only
  cases for which microscopically based derivations of elasticity are
  documented are (nearly) uniformly strained lattices. A microscopic approach
  to elasticity is proposed. As a first step, microscopically exact expressions
  for the displacement, strain and stress fields are derived. Conditions
  under which linear elastic constitutive relations hold are studied 
  theoretically and numerically.
  It turns out that  standard continuum
  elasticity is not self-evident, and applies only above certain spatial
  scales, which depend on details of the considered 
  system and  boundary conditions.
  Possible relevance to granular materials is briefly discussed.}


\maketitle
\section{Introduction} 
\label{sec:intro}
It is quite surprising that the existing microscopic justification of the
time-honored theory of elasticity, which has been thoroughly researched in a
variety of disciplines, is limited to lattice atomic
configurations~\cite{KittelFeynman}. Classical continuum elasticity theory has
been applied to a large variety of systems, including granular
materials~\cite{GranularElasticity}. In recent years the same theory has been
applied for the description of elastic properties of micro- and nano-scale
systems (e.g.,~\cite{MicroNano}). It is a-priori unclear whether this theory
applies at such small scales.

The study presented below shows that the justification of elastic theory based
on a microscopic picture is not entirely straightforward. As expected, one
finds that linear continuum elasticity is valid on sufficiently large scales.
Another result is that like in granular matter, one observes {\em force} chains
in strained elastic systems (also observed in~\cite{Wittmer01}). These chains
are not ``visible'' in the corresponding stress field. Classical mechanics is
assumed throughout this paper. The case of isostatic systems, which has
received considerable interest in the literature, is not specifically addressed
here.

\section{Coarse Graining and Constitutive Relations} 
\label{sec:cg}
\subsection{Preliminaries}
\label{sec:prelim}
Consider a system of particles (indexed by $\left\{i\right\}$) whose masses,
center of mass positions and velocities at time $t$ are given by $\{m_i;{\bf
  r}_i(t); {\bf v}_i(t)\}$.  Following~\cite{Glasser01} define the coarse
grained mass density at position $\vec{r}$ and time $t$ as
\begin{equation}
  \label{eq:cgmassdensity}
  \rho({\bf r},t)\equiv \sum_i m_i\phi[{\bf r}-{\bf r}_i(t)].
\end{equation}
Similarly, define the momentum density as
\begin{equation}
  \label{eq:cgmomentumdensity}
  \vec{p}({\bf r},t)\equiv \sum_i m_i \vec{v}_i(t) \phi[{\bf r}-{\bf
    r}_i(t)],
\end{equation}
where $\phi(\vec{R})$ is a normalized non-negative coarse graining function
(with a single maximum at $\vec{R}=0$) of width $w$, the coarse graining scale.
Unlike in \cite{Glasser01}, here only spatial, and not temporal
coarse graining, is invoked.  
Upon taking the time derivative of 
Eqs.~(\ref{eq:cgmassdensity},\ref{eq:cgmomentumdensity}) 
and performing straightforward algebraic
manipulations~\cite{Glasser01} one obtains two of the equations of continuum
mechanics. Eq.~(\ref{eq:cgmassdensity}) yields the equation of
continuity:
\begin{equation}
  \label{eq:continuity}
  \dot{\rho}({\bf r},t)= -\mbox{\boldmath{$\nabla$}}\cdot \vec{p}({\bf r},t)
  = -\mbox{\boldmath{$\nabla$}}\left[\rho({\bf r},t)\vec{V}({\bf r},t)\right] ,
\end{equation}
where $\dot{\rho}\equiv\pd{\rho}{t}$, and the coarse grained velocity field is
defined by \mbox{$\vec{V}({\bf r},t)\equiv \vec{p}({\bf r},t)/\rho({\bf
    r},t)$}. From Eq.~(\ref{eq:cgmomentumdensity}) one obtains the momentum
conservation equation:
\begin{equation}
  \label{eq:momentumconservation}
  \dot{p}_\alpha({\bf r},t)= -
  \frac{\partial}{\partial r_\beta} \left[ \rho({\bf r},t) V_{\alpha}({\bf
      r},t) V_{\beta}({\bf r},t) - \sigma_{\alpha\beta}({\bf r},t)\right],
\end{equation}
where Greek indices denote Cartesian coordinates. 

Define $f_{ij\alpha}(t)$ to be  the $\alpha$-th component of the force 
exerted on
particle $i$ by particle $j$ ($j \neq i$) at time $t$ (assuming pairwise
interactions), $\vec{r}_{ij}\equiv \vec{r}_{i}-\vec{r}_{j}$, and
the fluctuating velocity of particle $i$:
$\vec{v}'_{i}({\bf r},t)\equiv \vec{v}_i(t)-\vec{V}({\bf r},t)$.
With these definitions,
the stress tensor,
$\sigma_{\alpha\beta}$, is given by the following expression:
\begin{eqnarray}
\label{eq:stress}
\sigma_{\alpha\beta}({\bf r},t) &=&  -\frac{1}{2} \Bigg[ \sum_{i,j;i\ne j} 
f_{ij\alpha}(t) { r}_{ij\beta}(t)\\ 
&&\qquad\qquad\;\;\;\times \int_0^1 ds \phi[{\bf r}-{\bf r}_i(t) + 
s {\bf r}_{ij}(t)]\Bigg]\nonumber\\
&&-\sum_{i}m_i v'_{i\alpha}({\bf r},t) v'_{i\beta}({\bf r},t) 
\phi[{\bf r}-{\bf r}_i(t)], \nonumber
\end{eqnarray} 
The first term in Eq.~(\ref{eq:stress})
is commonly referred to as the ``contact stress'' (or ``collisional stress''), while the second term is a kinetic contribution (the ``kinetic
stress'' or ``streaming stress''), which vanishes for quasi-static
deformations.

The energy conservation equation can be obtained in a similar way. Assume, for
simplicity, that the forces are derived from a  potential function:
$\vec{f}_{ij}=-\mbox{\boldmath{$\nabla_i$}}
\Phi\left(\vec{r}_{ij}\right)$, with obvious notation.  Define the energy density as:
\begin{eqnarray}
  \label{eq:cgenergydensity}
  e({\bf r},t)&\equiv& \frac{1}{2} \sum_i m_i v^2_i(t) 
  \phi[{\bf r}-{\bf r}_i(t)] \nonumber\\
  &&+ \frac{1}{2} \sum_{i,j;i\ne j}  
  \Phi\left(\vec{r}_{ij}(t)\right) \phi[{\bf r}-{\bf r}_i(t)].
\end{eqnarray}
Application of a time derivative to $e({\bf r},t)$ and a rearrangement of terms
yields the energy conservation equation:
\begin{equation}
  \label{eq:energyconservation}
  \dot{e}({\bf r},t)= -
  \frac{\partial}{\partial r_\beta} \left[ V_{\beta}({\bf r},t)
    e({\bf r},t) - V_{\alpha}\sigma_{\alpha\beta}({\bf r},t) + c_{\beta}({\bf r},t)\right],
\end{equation}
where the heat flux, $\vec{c}$,  is given by:
\begin{eqnarray}
\label{eq:heatflux}
&&c_{\beta}({\bf r},t) =\\
&&  \frac{1}{2} \sum_{i} 
\left[ m_i v'^{2}_{i}({\bf r},t) + \sum_{j,j\neq i}
  \Phi\left(\vec{r}_{ij}(t)\right)
\right] v'_{i\beta}({\bf r},t)\phi[{\bf r}-{\bf r}_i(t)] \nonumber\\
&&+\frac{1}{4} \Bigg[ 
\sum_{i,j;i\ne j} \left[
v'_{i\alpha}({\bf r},t)+v'_{j\alpha}({\bf r},t)\right]
f_{ij\alpha}(t) {r}_{ij\beta}(t) \nonumber\\
&&\qquad\qquad\;\;\;\;\times \int_0^1 ds \phi[{\bf r}-{\bf r}_i(t) + s {\bf
  r}_{ij}(t)]\Bigg].\nonumber 
\end{eqnarray} 

\subsection{Displacement and Strain}
\label{sec:disp+strain}
Following elementary continuum mechanics, consider a {\em material particle}
whose initial (Lagrangian, at time $t=0$) coordinate is $\vec{R}$.  Its
(Eulerian) coordinate at time $t$ is denoted by $\vec{r(\vec{R},t)}$.  The
corresponding (Lagrangian) displacement field is given by \mbox{$\vec{u}^{\rm
    La}(\vec{R},t)\equiv \vec{r}(\vec{R},t)-\vec{R}$}. The material particle's
velocity is \mbox{$\vec{V}^{\rm La}(\vec{R},t)=\partial{\vec{u}^{\rm
      La}\left(\vec{R},t\right)}/\partial{t}$}. It therefore follows that
\mbox{$\vec{u}^{\rm La}(\vec{R},t)=\int_0^t \vec{V}^{\rm La}(\vec{R},t')dt'$}.
Using the definitions presented in Sec.~\ref{sec:prelim}, one obtains:
\begin{equation}
  \label{eq:displacement_la_exact}
\vec{u}^{\rm La}(\vec{R},t) \equiv
\int_0^t \frac{\sum_{i} m_i \vec{v}_i(t')
  \phi[\vec{r}(\vec{R},t')-\vec{r}_i(t')]} 
    {\sum_{j} m_j \phi[\vec{r}(\vec{R},t')-\vec{r}_j(t')]}dt'.
\end{equation}
The macroscopic displacement field $\vec{u}$ is history dependent, i.e., the
displacement at time $t$ depends on the trajectories of the particles from
$t=0$ to $t$. However, noting that $\dot{\vec{u}}_i=\vec{v}_i$, where
$\vec{u}_i\equiv\vec{r}_i(t)-\vec{r}_i(0)$ is the displacement of particle $i$,
and invoking integration by parts in Eq.~(\ref{eq:displacement_la_exact}), one
obtains (in the Eulerian representation):
\begin{eqnarray}
  \label{eq:displacement_eu_withcorrection}
  \vec{u}(\vec{r},t) &=& \vec{u}^\mathrm{lin}(\vec{r},t)\\
&&+\int_0^t dt' \frac{1}{\rho(\vec{r},t')}\pd{}{r_\beta}\bigg[
\sum_i m_i v'_{i\beta}(\vec{r},t') u'_{i\alpha}(\vec{r},t')\nonumber\\
&&\qquad\quad \times \phi[\vec{r}-\vec{r}_i(t')]\bigg],
\end{eqnarray}
where
\begin{equation}
  \label{eq:displacement_eu_linapprox}
  \vec{u}^\mathrm{lin}(\vec{r},t) \equiv
  \frac{\sum_{i} m_i \vec{u}_i(t)
    \phi[\vec{r}-\vec{r}_i(t)]} 
    {\sum_{j} m_j \phi[\vec{r}-\vec{r}_j(t)]},
\end{equation}
and
$\vec{u}'_{i}(\vec{r},t)\equiv\vec{u}_i(t)-\vec{u}^\mathrm{lin}(\vec{r},t)$.

It is claimed that $\vec{u}^\mathrm{lin}$ represents the displacement field
relevant to linear elasticity, i.e., the error is nonlinear in the strain. To
this end, let $\epsilon^{\star}_i=\max_j
\frac{\left|\vec{u}_i-\vec{u}_j\right|}{a_{ij}}$ where $j$ represents the
nearest neighbors of $i$ and $a_{ij}$ is the distance between the particles $i$
and $j$. Let $\epsilon^{\star}=\max_i \epsilon^{\star}_i$. Linear elasticity is
a theory which is linear in $\epsilon^{\star}$. Following the above
definitions,
\begin{equation}
  \label{eq:uiprime}
 u'_{i\alpha}= \frac{\sum_{j} m_i \left[{u}_{i\alpha}(t) -
    {u}_{j\alpha}(t)\right] \phi[\vec{r}-\vec{r}_i(t)]} {\sum_{k} m_k
  \phi[\vec{r}-\vec{r}_k(t)]},
\end{equation}
hence $\left|u'_{i\alpha}\right|<C_1\epsilon^{\star}$, where $C_1$ depends on
the coarse graining scale, and
$\left|v'_{i\alpha}\right|<C_2\epsilon^{\star}/\tau$, where $\tau$ represents
the typical time scale on which the $\left\{\vec{u}_i\right\}$ change (in the
quasistatic limit,\mbox{$\tau\rightarrow\infty$},  while $\epsilon^{\star}$ remains
finite).  The integrand in Eq.~(\ref{eq:displacement_eu_withcorrection}) is
thus $\mathcal{O}\left(\epsilon^{\star 2}\right)$. As a matter of fact, it is
easy to show that the integral on  the right hand side of
Eq.~(\ref{eq:displacement_eu_withcorrection}) is bounded from above by
$\mathcal{O}\left(\epsilon^{\star 2}\right)$. Note that
$\pd{\vec{u}'_{i}}{t}\neq\vec{v}'_{i}$, since $\vec{u}'_{i}$ is defined with
respect to the {\em linear} displacement $\vec{u}^\mathrm{lin}$ and not the
exact displacement, $\vec{u}$.  A useful feature of the linear displacement,
$\vec{u}^\mathrm{lin}(\vec{r},t)$, is that, unlike the exact displacement field
[Eq.~(\ref{eq:displacement_eu_withcorrection})], it depends only on the
microscopic displacements at time $t$.

The linear strain field is \mbox{$\epsilon_{\alpha\beta}(\vec{r},t)=
  \frac{1}{2}\left[\frac{\partial u_\alpha(\vec{r},t)}{\partial r_\beta} +
    \frac{\partial u_\beta(\vec{r},t)}{\partial r_\alpha}\right]$}. Notice that
$\epsilon_{\alpha\beta}=\mathcal{O}\left(\epsilon^{\star}\right)$. Following
the above arguments, this expression can be replaced by:
\begin{equation}
  \label{eq:linstrain}
\epsilon^\mathrm{lin}_{\alpha\beta}(\vec{r},t)=
  \frac{1}{2}\left[\frac{\partial u^\mathrm{lin}_\alpha(\vec{r},t)}{\partial r_\beta} +
    \frac{\partial u^\mathrm{lin}_\beta(\vec{r},t)}{\partial r_\alpha}\right],
\end{equation}
the error being $\mathcal{O}\left(\epsilon^{\star 2}\right)$.

\begin{figure}[!ht]
  \begin{center}
    \includegraphics[width=3.5in]{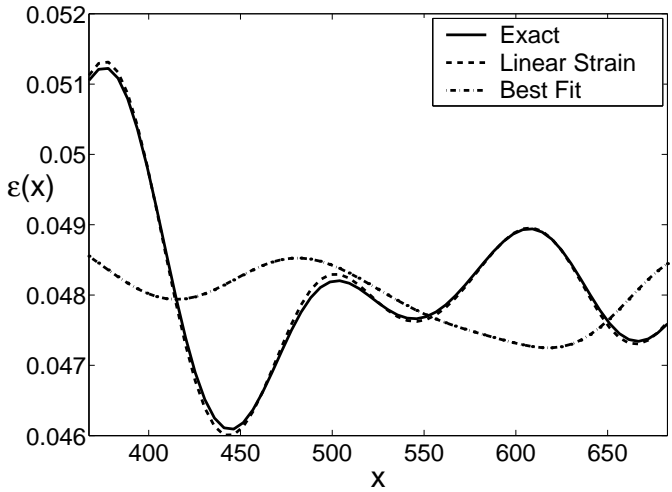}
  \end{center}
  \caption{The Eulerian strain $\epsilon(x)$ vs. position $x$ in the
    central region of a linear chain of particles connected by linear springs
    of random  spring constants and  rest lengths, calculated by three
    methods (see text).}
  \label{fig:StrainComparison1D}
\end{figure}

It is interesting to compare the above results with some previously defined
heuristic calculations  of
the strain field. The mean field strain  (e.g.~\cite{Bathurst88}) is based
on the assumption that the relative particle displacements are described by the
macroscopic strain, i.e., \mbox{$u_{ij\alpha}(\vec{r},t)=\epsilon_{\alpha\beta}
  r_{ij\beta}$} where $\vec{u}_{ij}\equiv \vec{u}_{i}-\vec{u}_{j}$ (an affine
deformation). An  improvement of this method, 
which enables a local evaluation of the
strain field, is provided by the ``best fit'' hypothesis~\cite{Liao97}, 
whereby the rms difference between the actual relative displacements
and the above mean field expression for them is minimized in
a given volume,  to produce a `best' strain field.
The mean field approaches are in
general inconsistent with local force equilibrium (except for homogeneous
deformations of lattice configurations).  Therefore, the mean field (or best
fit) strain constitutes  an uncontrolled approximation of the strain field. The
difference between the exact strain field, the linear strain as given by
Eq.~(\ref{eq:linstrain}), and the best fit approximation are demonstrated in
Fig.~\ref{fig:StrainComparison1D}. In this figure, the above three fields are
presented for a one dimensional chain of $1000$ point 
particles connected by linear
springs of random rest lengths (mean rest length $a=1$, relative standard
deviation $0.29$) and random spring constants (relative standard deviation
$0.13$), the global  strain of the system being $\epsilon=0.05$.  
To this end, a Gaussian
coarse graining function $\phi(\mathbf{r})=\frac{1}{\pi
  w^2}e^{-(|\mathbf{r}|/w)^2}$ with \mbox{$w=50a$} has been employed. In the
calculation of the best fit strain, the fluctuations with respect to the mean
field are weighted using the same coarse graining function as in the exact
formulation. As seen in Fig.~\ref{fig:StrainComparison1D}, the linear strain is
very close to the exact strain, whereas the best fit provides quite a poor
approximation (for the case of equal spring constants, the best fit method
yields a uniform strain, while the correct strain is space
dependent~\cite{GoldenbergUP}). In general, heuristic approximations for the
strain field may result in inaccurate constitutive relations.
\subsection{Stress-Strain Relation}
\label{sec:stress_strain}
It is not a-priori clear that the stress field can be expressed as a linear
functional of the strain field, even for small deformations. Each of these two
macroscopic fields, cf. Eqs.~(\ref{eq:stress},\ref{eq:linstrain}), is a {\em
  different} average of microscopic entities. Once averaging is invoked to
produce macroscopic fields, on cannot deduce the microscopic entities from
these fields. This is one of the fundamental flaws of the mean field
approaches, which rely on such a deduction. Below, an exact method for
obtaining linear elasticity, which demonstrates the above problem and
highlights the scale limitations of this theory, is outlined.

Consider, for sake of simplicity, only systems with pairwise interactions. In
order to develop linear elasticity one can assume, without loss of generality,
harmonic interactions:
\begin{equation}
  \label{eq:harmonicpotential}
 \Phi\left(\vec{r}_{ij}(t)\right)=
\frac{1}{2}K_{ij}\left(|\vec{r}_{ij}|-l_{ij}\right)^2,
\end{equation}
where $l_{ij}$ is the equilibrium separation of particles $i$ and $j$. The
force on particle $i$ exerted by particle $j$ is given, to linear order in the
relative particle displacements,  $\vec{u}_{ij}$, by:
\begin{equation}
  \label{eq:harmonicforceslinear}
  \vec{f}_{ij}\simeq-K_{ij}\left(\hat{\vec{r}}^0_{ij}\cdot \vec{u}_{ij}
  \right) \hat{\vec{r}}^0_{ij},
\end{equation}
where the superscript~$0$ denotes the reference configuration, in which all
particle pairs are at their equilibrium separation
($\left|\vec{r}^0_{ij}\right|=l_{ij}$), i.e., an unstressed configuration
(prestressed states  are not considered here).

It is apparent that even in this case, the microscopic expressions for the
contact stress [the first term in Eq.~(\ref{eq:stress})] and the strain
[Eq.~(\ref{eq:linstrain})] are not manifestly proportional.  Therefore
macroscopic elasticity is not a-priori obvious. To see how it still comes
about, substitute Eq.~(\ref{eq:harmonicforceslinear}) in Eq.~(\ref{eq:stress}).
The contact stress, to linear order in $\left\{\vec{u}_{ij}\right\}$), is:
\begin{eqnarray}
  \label{eq:lin_stress}
 \sigma_{\alpha\beta}^\mathrm{lin}({\bf r},t) &=& \frac{1}{2} \Bigg[
  \sum_{ij} K_{ij}\hat{r}^0_{ij\gamma} u_{ij\gamma}\hat{r}^0_{ij\alpha}
  {r}^0_{ij\beta}\\
&&\quad\;\;\;\times \int_0^1 ds \phi[{\bf r}-{\bf r}^0_i + 
s {\bf r}^0_{ij}]\Bigg].\nonumber
\end{eqnarray}

Consider a volume $\Omega$, whose linear dimension, $W$,
is much larger than the coarse graining scale,
$w$, and let $\vec{r}$ be an interior point of $\Omega$ which is `far'
from its boundary.  Let upper case Latin indices denote the particles in the
exterior of $\Omega$ which interact with particles inside $\Omega$.  Since the
considered system is linear, there exists a Green's function $\bf G$ such that
\mbox{$u_{i\alpha}=G_{i\alpha J\beta}u_{J\beta}$} for $i\in \Omega$.
Let: \mbox{$L_{ij\alpha J\beta}\equiv G_{i\alpha J\beta}-G_{j\alpha  J\beta}$}. It follows that:
\mbox{$u_{ij\alpha}= L_{ij\alpha J\beta} u_{J\beta}$}.
Under a rigid translation (all \mbox{\{ $\vec{u}_J \}$} equal): \mbox{$\vec{u}_{ij}=0$}, hence:
\mbox{$u_{ij\alpha}= L_{ij\alpha J\beta}
\left[u_{J\beta}-u_\beta(\vec{r})\right]$}. It follows that
\begin{equation}
  \label{eq:relativedisp}
  u_{ij\alpha}=L_{ij\alpha J\beta}
  \left[u_\beta(\vec{r}_J)-u_\beta(\vec{r})\right] +L_{ij\alpha J\beta}
  \left[u_{J\beta}-u_\beta(\vec{r}_J)\right],
\end{equation}
where \mbox{$u_{J\beta}-u_\beta(\vec{r}_J)$} is a fluctuating displacement. The
sum over $J$ in the second term can be shown to be subdominant when  $W$
sufficiently exceeds $w$.
The first term equals, to leading  order in a gradient expansion:
\begin{equation}
  \label{eq:gradient_expansion}
u_\beta(\vec{r}_J)-u_\beta(\vec{r})\simeq
  \pd{u_{\beta}(\vec{r})}{r_{\gamma}}\left(r_{J\gamma}-r_\gamma\right).  
\end{equation}
Substituting Eq.~(\ref{eq:gradient_expansion}) in Eq.~(\ref{eq:lin_stress}):
\begin{eqnarray}
\label{stress_strain_linear}
\sigma_{\alpha\beta}(\vec{r}) &\simeq&  \frac{1}{2}
\Bigg[ \sum_{ij} K_{ij} L_{ij\gamma J\mu}
\left(r^0_{J\nu}-r_\nu\right)
\hat{r}^0_{ij\alpha}{r}^0_{ij\beta}  \hat{r}^0_{ij\gamma} 
\\
&& \qquad \times \int_0^1 ds \phi[{\bf r}-{\bf r}^0_i + 
s {\bf r}^0_{ij}]\Bigg] \epsilon_{\mu\nu}(\vec{r}),\nonumber
\end{eqnarray} 
where rotational symmetry has been invoked.
Thus linear elasticity is valid when
$\left\|\mbox{\boldmath{$\epsilon$}}\right\|\ll 1$ (the strain components are
small) and $\left| W \nabla_{\alpha} \nabla_{\beta} \vec{u} \right|\ll 1$. Note
that the elastic moduli depend, in principle, on the position as well as the
resolution (through the coarse graining function $\phi$). Our numerical results
(see below) indicate that the contribution of the second term on the right hand
side of Eq.~(\ref{eq:relativedisp}) to the stress is smaller than a naive bound
based on surface to volume ratios implies.

\subsection{Elastic Energy}
\label{sec:elastic_energy}
In the quasistatic limit, the energy density  reduces, cf.
Eq.~(\ref{eq:cgenergydensity}), to:
\begin{equation}
  \label{eq:energydensity_novelocity}
  e({\bf r},t)=\frac{1}{2} \sum_{i,j;i\ne j}  
  \Phi\left(\vec{r}_{ij}(t)\right) \phi[{\bf r}-{\bf r}_i(t)].
\end{equation}
To lowest nonvanishing order in the strain, the potential energy corresponding
to Eq.~(\ref{eq:harmonicpotential})  is given by
$\frac{1}{2}K_{ij}\left(\hat{\vec{r}}^0_{ij}\cdot \vec{u}_{ij} \right)^2$.
Hence, at this order:
\begin{equation}
  \label{eq:elasticenergydensity}
  e({\bf r},t)= \frac{1}{4} \sum_{i,j;i\ne j}  
K_{ij}\left(\hat{\vec{r}}^0_{ij}\cdot \vec{u}_{ij} \right)^2
\phi[{\bf r}-{\bf r}_i(t)].
\end{equation}
If linear elasticity is to hold, one must have:
$e=\frac{1}{2}\mbox{\boldmath{$\sigma$}}\cdot\mbox{\boldmath{$\epsilon$}}$. A
set of straightforward transformations on Eq.~(\ref{eq:elasticenergydensity})
yields:
\begin{eqnarray}
  \label{eq:energy_equation_quasistatic}
  e({\bf r},t)&=&
\frac{1}{2}\mbox{\boldmath{$\sigma$}}\cdot\mbox{\boldmath{$\epsilon$}}
- \pd{}{r_{\beta}}  \Bigg[ \frac{1}{4}
\sum_{i,j;i\ne j}  \left(\vec{f}_{ij}\cdot \vec{u}'_i\right) r_{ij\beta}\\
&& \qquad\qquad\qquad\;\;
\times \int_0^1 ds \phi[{\bf r}-{\bf r}_i(t) + s {\bf
  r}_{ij}(t)]\Bigg].\nonumber 
\end{eqnarray}
It can be shown that the second term on the right hand side of
Eq.~(\ref{eq:energy_equation_quasistatic}) represents the adiabatic limit of
the divergence of the heat flux [Eq.~(\ref{eq:heatflux})], i.e., the work of
the fluctuating forces at the surface of a control volume. As this term is
a  divergence of a flux, its average over a sufficiently large
volume tends to zero, i.e.  the {\em average} of
$\frac{1}{2}\mbox{\boldmath{$\sigma$}}\cdot\mbox{\boldmath{$\epsilon$}}$ over a
sufficiently large volume (not its `local value') 
is  the elastic energy density. As in the
previous section, one obtains that classical elasticity is valid only for
sufficiently large scales, in particular scales for which ``surface
contributions'', as explained above, vanish.

\section{Numerical Results}
\label{sec:numerical}
The above results are demonstrated on a  two dimensional (2D)
system of particles with harmonic interactions
[Eq.~(\ref{eq:harmonicforceslinear})]. As a first test case, consider a
square-shaped triangular lattice configuration, with uniform nearest-neighbor
spring constants $K$ and rest lengths equal to the lattice constant $d$,
subjected to the following boundary conditions. The displacements of the
particles at the boundary (whose positions are denoted by
$\left\{\vec{r}^0_I\right\}$) are chosen to yield a homogeneous deformation,
i.e., the boundary is subject to an ``applied strain'',
$\epsilon_{\alpha\beta}$:
\begin{equation}
  \label{eq:starin_bc}
u_{I\alpha}=\epsilon_{\alpha\beta}\left(r_{I\beta}-r_{0\beta}\right),  
\end{equation}
where $r_{0\beta}$ is an arbitrary point, chosen to reside at a corner of the
system.  The (linear) static equilibrium equations are solved (by matrix
inversion) for a given applied $\epsilon_{\alpha\beta}$, yielding (for each 
choice of $\epsilon_{\alpha\beta}$)
a set of displacements $\left\{\vec{u}^{\mbox{$\boldmath{\epsilon}$}  }_i\right\}$. The latter are used for
calculating the linear strain field [using Eq.~(\ref{eq:linstrain}] and 
linear stress, Eq.~(\ref{eq:lin_stress}). The coarse-graining function used is,
\begin{equation}
  \label{eq:fermi_cg}
  \phi(\vec{R})=\phi(R)=\frac{A}{1+e^\frac{R-w/2}{\lambda}}
\end{equation}
(the Fermi distribution), chosen as a smoothed Heaviside function. The length
on which the function decays to zero, $\lambda$, can be chosen independently of
the coarse graining scale,  $w$.  In all the calculations presented below,
$\lambda=\frac{d}{2}$. The constant $A$ is fixed by the normalization: in 2D,
$2\pi \int_0^\infty R \phi(R) dR=1$.

In this case of an ordered lattice, the strain components converge to the
applied strain components even for $w=d$. This should be expected, since, as
mentioned, for a lattice configuration (with uniform spring constants) under
homogeneous deformation, the particle displacements correspond to an affine
transformation, rendering the mean field approximation exact. The corresponding
stress components are  scale-independent as well,   and consistent with the
continuum isotropic elastic moduli for a triangular lattice (Lam{\'e} constants
$\lambda=\mu=\frac{\sqrt{3}K}{4}$).  

For disordered systems, even for a ``homogeneous applied strain'' as described
above, the stress and strain fields are inhomogeneous, in general. Linear
elasticity should still be valid (at least as an approximation, as described in
Sec.~\ref{sec:cg}),  i.e. the local stress should depend on  the local
strain by an appropriate local  linear relation on a sufficiently large
coarse graining scale.  In order to examine the influence of disorder on
the validity of linear elasticity, disordered systems were generated, based on
the triangular lattice configuration: a random number, uniformly distributed in
the range $[-\delta d,\delta d]$, is added to the $x$ and $y$ coordinates of
the particles. Particles whose distance is less than $c_\mathrm{max}=1.1d$ are
connected by springs, with rest lengths equal to the particle separation in the
initial configuration (ensuring an unstressed configuration), and spring
constants uniformly distributed in the range $[K-\delta K,K+\delta K]$. Note
that this choice of $c_\mathrm{max}$ can decrease the coordination number, for
some of the particles, to a value smaller than  6, for sufficiently large
$\delta d$ (this kind of topological disorder can give rise to qualitatively
different effects than positional disorder). To evaluate the extent to which
disordered systems are described by the equations of linear elasticity, the
following procedure is used. Three independent global strains
$\epsilon^i_{\alpha\beta};\ i=1,2,3$ are applied, and the stress and strain
fields are calculated at a given point. According to linear elasticity, these
fields should be linearly related, though the elastic moduli may be position
dependent. In a 2D system with central forces (as used here), there are, in
general, 6 independent elastic moduli. Each deformation provides three linear
equations for these moduli (however, it can be shown that two independent
deformations are insufficient for determining them, and three deformations are
required). The elastic moduli are determined using 6 of the 9 linear equations.
The stress components which are not used in this procedure are then calculated
using these elastic moduli, and their values are compared to those computed
directly using Eq.~(\ref{eq:lin_stress}). The root mean square
of the differences between the  stress components calculated by employing
the measured moduli and their directly measured exact  values
(normalized by the norm of the  exact values),
$\Delta$, is used as a measure of the extent to which the system
is described by linear elasticity at a given position and for a given value
of the coarse graining
scale.

\begin{figure}[!ht]
  \begin{center}
    \includegraphics[width=3.5in]{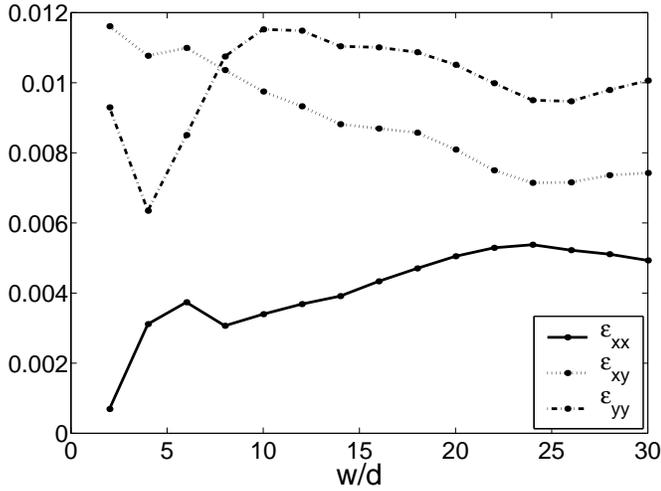}
  \end{center}
  \caption{The strain components vs. the coarse graining scale, $w$, in a disordered
    configuration (see text), for an applied strain
    $\mbox{\boldmath{$\epsilon$}}^1$ (see text).} 
  \label{fig:strain_disordered}
\end{figure}

\begin{figure}[!ht]
  \begin{center}
    \includegraphics[width=3.5in]{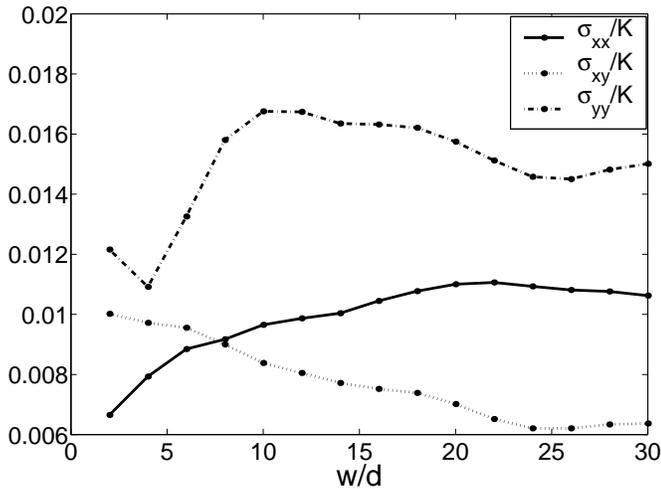}
  \end{center}
  \caption{The stress components vs. the coarse graining scale, $w$, in a disordered
    configuration (see text), for an applied strain
    $\mbox{\boldmath{$\epsilon$}}^1$ (see text).}
  \label{fig:stress_disordered}
\end{figure}

The stress and strain components at the center of a disordered system of size
$40d\times40d$, with $\delta d=0.1d$, \mbox{$\delta K=0.1K$} and applied strain
$\mbox{\boldmath{$\epsilon$}}^1=\left(\protect\begin{array}{cc} 0.005 & 0.0075\protect\\
    0.0075 & 0.01\protect\end{array}\right)$, \,  are shown in
Figs.~\ref{fig:strain_disordered},\ref{fig:stress_disordered}, respectively,
for different coarse graining scales.  These fields are obviously inhomogeneous, hence the observed scale dependence (note that for coarse
graining scales approaching the system size, the values of the strain
components do approach their  imposed global values, as
expected).


\begin{figure}[!ht]
  \begin{center}
    \includegraphics[width=3.5in]{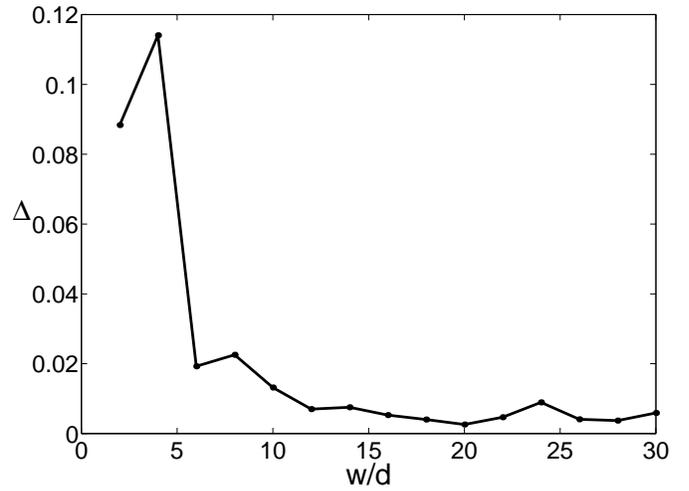}
  \end{center}
  \caption{The deviation from elasticity in a disordered
    configuration (see text), measured by $\Delta$ (see text), vs. the coarse
    graining scale, $w$.}
  \label{fig:elastic_dev_disordered}
\end{figure}

\begin{figure}[!ht]
  \begin{center}
    \includegraphics[width=3.5in]{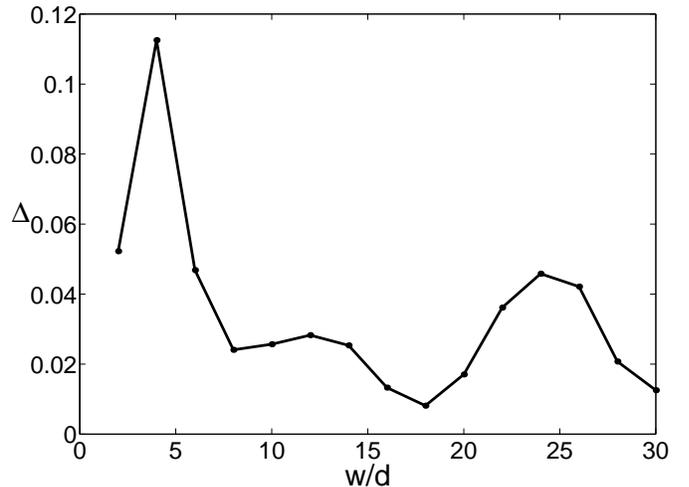}
  \end{center}
  \caption{The deviation from elasticity in a 
    configuration with higher disorder compared to
    Fig.~\ref{fig:elastic_dev_disordered} (see text), measured by $\Delta$ (see
    text), vs. the coarse graining scale, $w$.}
  \label{fig:elastic_dev_more_disordered}
\end{figure}

Fig.~\ref{fig:elastic_dev_disordered} shows the deviations from elasticity,
measured by $\Delta$ as defined above, for the same disordered system. The
deviation is quite large for small coarse graining scales, but it decreases to
less than 1\% for $w>12d$, indicating that linear elasticity holds reasonably
well beyond this scale. Note that at this scale, the stress and strain are
still scale dependent, or inhomogeneous, which implies that linear elasticity
does hold {\em locally} inside the system, with elastic moduli which depend on
position (and  scale). The deviations from elasticity for a similar
system with higher disorder (with $\delta d=0.15d$, $\delta K=0.15K$) is shown
in Fig.~\ref{fig:elastic_dev_more_disordered}.
One may
interpret the observed fluctuations as an indication that 
the scale required to obtain linear elasticity exceeds the size of the
system, as even at scales close to this size there is no clear  saturation
to a linear elastic relation.

\section{Concluding Remarks}
\label{sec:conclusions}
It is important to note that the results presented in Sec.~\ref{sec:numerical}
have been obtained with boundary conditions chosen to obtain a homogeneous
strain field.  Indeed, in the case of a lattice, a uniform strain 
is obtained. Thus the inhomogeneity in the disordered case is not a result
of the applied boundary conditions but of the disorder itself.
When inhomogeneous  boundary conditions are applied one expects
an inhomogeneous strain field even  in a homogeneous system. In this
case one should observe  deviations from
linear elasticity on small scales 
or close to the boundary.
The deviations from standard linear elasticity should be particularly prominent in small
systems. This is indeed the case~\cite{Goldenberg01} for 
an ordered (lattice configuration) slab of particles 
resting on a rigid support (``bottom'') with a point force applied to
the center particle at the ``top'' (motivated by experiments on granular
systems~\cite{Clement,Geng}).  A comparison of the stress field obtained at the
bottom of slabs of different heights (number of layers of particles) with 
corresponding continuum elastic solutions~\cite{Goldenberg01} shows significant deviations,
for a small number of layers. These deviations decrease as the size
of the system increases,
rendering continuum elasticity a good approximation
for a sufficient number of layers [$\mathcal{O}(40)$ in 2D,
$\mathcal{O}(60)$ in 3D].  For disordered systems, this effect is  even
more pronounced, as elasticity sets in on larger coarse graining scales (which
should be compared to the size of the system). An
additional factor  which is expected to influence the crossover to
linear elasticity is a possible 
 inhomogeneous stress in the reference state, as
observed in~\cite{Wittmer01} (in the examples presented here, the reference
state is unstressed).  In~\cite{Wittmer01}, a similar crossover has been
observed for the {\em vibrational modes} of disordered systems, in which the
strain is typically inhomogeneous. While not mentioned in~\cite{Wittmer01},
their results appear to suggest that the crossover is obtained for larger
system sizes as the frequency (wavelength) is increased, i.e., as the strain
gradients are larger,  consistently  with the above arguments.

The above considerations, as well as the theoretical calculations described in
Sec.~\ref{sec:cg}, suggest that the general constitutive description of the
systems considered here, even for small deformations, should involve a
non-local relation between the stress and the strain field, a leading
approximation thereof being provided by {\em gradient elasticity}. 
Larger deformations should
obviously require nonlinear elasticity.

While in granular matter, the interactions among the grains are not harmonic,
we believe that the above discussion may still be relevant for quasi-static
deformation of granular materials. First, the interactions are often described
by elastic contact models for which the force-displacement law, even when
nonlinear, may be linearized around a reference state (though to conform with
the nature of cohesionless grains, the springs should be ``one-sided'', i.e.,
allow for compressive forces only). A deformation under which the contact
network does not change should then still be described by linear elasticity on
sufficiently large scales. As the interactions among the grains 
are only compressive, some
contacts may break under a given applied boundary conditions, yielding a
modified contact network.  For small deformations, the changes in this network
may be sufficiently small for the elastic moduli not to be affected. When this
is not the case, {\em incremental} elasticity, whereby the elastic moduli are
modified by the deformation, may still be appropriate for describing the
deformation. Furthermore,  when
the boundary conditions result in tensile stress components in a given region,
contacts may break there preferentially in specific
directions~\cite{Goldenberg01,Luding97}. A similar type of stress-induced
anisotropy has been suggested in the context of plastic models for soil
mechanics~\cite{Oda}.  Another source of deviations from elasticity is granular
friction.  Note, however, that static friction is not dissipative,
and it may actually prevent the breaking of contacts, extending the elastic
range of frictional systems with respect to the (idealized) frictionless case.
Once the limit of static friction is exceeded, friction is kinetic and
dissipation does occur.  In this case one expects plastic failure, which is
clearly beyond the limits of validity of elastic theory.

As a final remark, force chains, which have been observed in both
experiments~\cite{ForceChainsExp} and simulations~\cite{ForceChainsTheory} of
granular materials, are also observed in elastic
systems~\cite{Wittmer01,Goldenberg01}. However, as shown
in~\cite{Goldenberg01}, force chains appear even in inhomogeneously strained
lattices which are macroscopically isotropic, due to the inherent small-scale
anisotropy of discrete systems. The corresponding stress field, even at small
scales, does not exhibit similar structures in an isotropic system, i.e., force
chains do not necessarily imply an inhomogeneous stress field.

\begin{acknowledgement}
\label{sec:ack}
Support from the Israel Science Foundation, grants no. 39/98 and 53/01, is
gratefully acknowledged.
\end{acknowledgement}


\begin{thebibliography}{10}
  
\bibitem{KittelFeynman} C.~Kittel, \newblock {\em Introduction to Solid State
    Physics (Second Edition)} (Wiley, 1956); R.~P. Feynman, R.~B. Leighton, and
  M.~Sands, \newblock {\em The Feynman Lectures on Physics, Vol. II}
  (Addison-Wesley, 1964); M. Born and K. Huang, {\em Dynamical Theory of
    Crystal Lattices} (Clarendon Press, 1954).
  
\bibitem{GranularElasticity} R.~M.  Nedderman, \newblock {\em Statics and
    Kinematics of Granular Materials} (Cambridge University Press, 1992); S.~B.
  Savage, \newblock in {\em Proc. NATO ASI on Physics of Dry Granular Media,
    Carg{\`e}se, France, 1997}, edited by H.~J.  Herrmann, J.~P. Hovi, and
  S.~Luding, pp.  25--95, Kluwer, 1998.

\bibitem{MicroNano} J.~Schi{\o}tz et al., \newblock Phys. Rev. B {\bf 60},
  11971 (1999); A.~G.  Khachaturyan, S.~Semenovskaya, and T.~Tsakalakos,
  \newblock Phys. Rev. B {\bf 52}, 15909 (1995); R.~Lifshitz and M.~L. Roukes,
  \newblock Phys.  Rev. B {\bf 60}, 5600 (2000).
  
\bibitem{Wittmer01} J.~P.~Wittmer et al., cond-mat/0104509.
  
\bibitem{Glasser01}
B.~J. Glasser and I.~Goldhirsch,
\newblock Physics of Fluids {\bf 13}, 407 (2001).

\bibitem{Bathurst88} R.~J. Bathurst and L.~Rothenburg, \newblock J. Appl.
  Mech. {\bf 55}, 17 (1988).

\bibitem{Liao97} C.-L. Liao, T.-P. Chang, D.-H. Young, and C.~S. Chang,
  \newblock Int. J. Solids and Structures {\bf 34}, 4087 (1997).
  
\bibitem{GoldenbergUP} C.~Goldenberg and I.~Goldhirsch, \newblock Unpublished.

\bibitem{Goldenberg01} C.~Goldenberg and I.~Goldhirsch, cond-mat/0108297.
    
\bibitem{Clement} G.~Reydellet and E.~Cl{\'e}ment, \newblock Phys. Rev.  Lett.
  {\bf 86}, 3308 (2001); D.~Serero et al.,  \newblock Eur. Phys. J. E {\bf 6},
  169 (2001).

\bibitem{Geng} J.~Geng {\em et~al.}, \newblock Phys. Rev.  Lett. {\bf 87},
  035506 (2000).
  
\bibitem{Luding97} S.~Luding, \newblock Phys. Rev. E {\bf 55}, 4720 (1997).

\bibitem{Oda} M. Oda, \newblock Mech. Mat. {\bf 16}, 35 (1993).

\bibitem{ForceChainsExp} A.~Drescher and G.~de~Josselin~de Jong, \newblock J.
  Mech. Phys. Solids {\bf 20}, 337 (1972).
  
\bibitem{ForceChainsTheory} P.~A. Cundall and O.~D.~L. Strack, \newblock
  Geotechnique {\bf 29}, 47 (1979); F.~Radjai, S.~Roux, and J.~J. Moreau,
  \newblock Chaos {\bf 9}, 544 (1999).

\end{thebibliography}
\end{document}